Article type:   Original research

Title:   Gravitational anti-screening as an alternative to dark matter

Author:   A. Raymond Penner

Address:   Department of Physics,
Vancouver Island University,
900 Fifth Street,
Nanaimo, BC, Canada,
V9R 5S5

Email:   raymond.penner@viu.ca

Tel:   250 753-3245 ext: 2336

Fax:   250 740-6482

# Gravitational anti-screening as an alternative to dark matter

## Abstract

A semiclassical model of the screening of electric charge by virtual electric dipoles, as found in electrodynamic theory, will be presented. This model is then applied to the hypothetical case of an electric force where like charges attract. The resulting anti-screening of the electric charge is found to have the same functional dependence on the field source and observation distance that is found with the Baryonic Tully-Fisher Relationship. This leads to an anti-screening model for the gravitational force which is then used to determine the theoretical rotational curve of the Galaxy and the theoretical velocity dispersions and shear values for the Coma cluster. These theoretical results are found to be in good agreement with the corresponding astronomical observations. The screening of electric charge as found in QED and the larger apparent masses of galaxies and galactic clusters therefore appears to be two sides of the same coin. **Key words:** cosmology: dark matter – general: gravitation

## 1. Introduction

Dark matter is a hypothetical type of matter that does not absorb or emit radiation at any significant level. It is generally believed to be a subatomic particle with the prime candidate being stable weakly interacting massive particles (WIMPs). Although dark matter is evoked as an explanation for a variety of observations the hypothesis stems primarily from the following two. First, the rotational velocities of stars and gas clouds in the outer regions of spiral galaxies are found to be much greater that what is predicted from the baryonic mass of these galaxies. Examples of galactic rotational curves are provided by Sofue (1996), Sofue et al (1999) and Nordermeer et al (2007). In general, the rotational velocities are found to approach a constant value in the outer regions with little indication that they will eventually fall off in a Keplerian fashion. It is difficult to determine exactly how much additional mass is needed. However, it would appear that an order of magnitude more mass than what is provided by the baryonic mass is required. The second set of observations that led to the dark matter hypothesis is that the velocity dispersions of galaxies in galactic clusters also indicate that approximately an order of magnitude more mass than what is provided by the baryonic mass is required (Lokas&Mamon 2003, Merritt 1987). Indeed, this result is what first led to the idea of dark matter (Zwicky 1933). The additional mass required by clusters is confirmed by shear measurements (Kubo et al 2007, Gavazzi et al 2009) and by observations of the X-ray emission of the intra-cluster gas (Hughes 1989, Briel et al 1992) which is the dominant baryonic component of clusters.

Although, the dark matter hypothesis does provide an explanation for these two sets of observations it is far from satisfactory. In the opinion of this author, there are several major problems. First, no WIMPs have been detected. If future experiments with the LHC come up empty it should be concluded that no such particles exist. Second, the modeling of the expected dark matter distribution leads to a profile (Navarro et al 1996, 1997) that is at best only moderately successful in matching the observed rotational curves of spiral galaxies even with the distributions free parameters. Specifically, it does not naturally lead to the rotational curves approaching a constant value in the outer regions. Finally, there is a primary observation for which the standard dark matter theory does not provide a natural explanation, namely the Baryonic Tully-Fisher relationship (BTFR) (McGaugh et al 2000, McGaugh 2012). The BTFR is the empirical relationship that exists between $M_o$, the baryonic mass of a galaxy, and v, the galaxy's constant outer rotational velocity. The BTFR as given by McGaugh (2012) is

$$M_o = A\, v^4 \quad \text{with} \tag{1a}$$

$$A = (47 \pm 6)\, M_\odot\, \text{km}^{-4}\, \text{s}^4 = (9.4 \pm 1.2) \times 10^{19}\, \text{kg}\, \text{m}^{-4}\, \text{s}^4. \tag{1b}$$

The BTFR can also be expressed as the following relationship between M, the apparent mass of the galaxy, its baryonic mass, and the observation distance r, by substituting GM/r for $v^2$ in (1a);

$$M = (G^2 A)^{-1/2} r\, M_o^{1/2}. \tag{2}$$

As (2) shows, in the outer regions of a galaxy the apparent mass increases linearly with observation distance and depends on the square root of the baryonic mass. The BTFR is a relationship with surprisingly little scatter that ranges over five orders of magnitude of galactic baryonic masses.

It is the inability of the standard dark matter model to naturally explain the BTFR that has been a primary motivation behind alternatives to the dark matter theory. For example, a leading alternative is MOdified Newtonian Dynamics (MOND) as proposed by Milgrom (1983, 1994, and 2002). In MOND it is postulated that the inertia of an object varies with acceleration in a manner that specifically produces the BTFR.

Another, less well known, alternative to dark matter is based on the hypothesis that mass dipoles exist throughout the cosmos. These dipoles become gravitationally polarized in the presence of an external gravitational field which leads to an anti-screening of the baryonic mass. Several models have previously been put forth based on this hypothesis. These models differ in what the mass dipoles actually are and how they behave in a gravitational field.

Blanchet (2007a) hypothesized a mass dipole wherein one particle has positive gravitational mass while the other has negative gravitational mass and with both having positive inertial masses. An external gravitational field is taken to exert a torque on the dipole moment causing the dipoles to align with the gravitational field. As the author indicates this particular model violates the equivalence principle. Blanchet (2007b) and Blanchet and Le Tiec (2008) followed up with a model in which a hypothesized particle has a passive gravitational mass equal to its inertial mass, in agreement with GR. However, the active gravitational mass is not equal to its inertial mass. The monopole component of the active mass is taken to be negligible but the particle has an active gravitational dipole moment which is what provides the contribution to the gravitational field. An issue with this model is that having the passive gravitational mass positive and equal to the inertial mass but not equal to the active gravitational mass would seem to violate Newton's 3rd law and thereby the conservation of momentum.

Another approach to the polarization has been undertaken by Hajdukovic (2011a, 2011b, 2011c, 2012a, 2012b). Hajdukovic's hypothesis is based on the idea that antiparticles have negative gravitational mass but a positive inertial mass and will be gravitationally repelled by normal matter. Virtual particle-antiparticle pairs that arise within the vacuum are taken then to align in an external gravitational field. The specific particle-antiparticle pairs considered by Hajdukovic are virtual pions. As with Blanchet's initial model, having the inertial mass positive and the gravitational mass negative for a particle violates the equivalence principle. In addition, the properties attributed to the antiparticles do not agree with relativistic quantum theory where an antiparticle has both a positive mass and a positive energy.

Penner (2012) also hypothesized that virtual particle-antiparticle pairs with equal and opposite energy or gravitational mass but positive inertial mass are the source of gravitational polarization. In this model the virtual pair was taken to be unbound. This model suffers from the same major defects as with Hajdukovic's model and the author no longer considers this a suitable model. However, the application of this mass dipole model to the determination of the rotational curve for the Galaxy (Penner 2013a), the general behaviour of rotational curves for spiral galaxies (Penner 2013b), and the dynamics of the Coma cluster (Penner 2013c) did show that a theory of the gravitational polarization of the vacuum agrees with several observations.

Separate from these attempts to evoke a mass dipole field as an alternative to dark matter, Bondi (1957) found a solution to Einstein's GR equations which does allow for a mass dipole consisting of two particles of equal and opposite mass. In Bondi's solution all three of each of the individual particle masses, i.e. inertial, passive gravitational, and active gravitational, are equal. If the pair is bound non-gravitationally it is found to behave in a strange manner in that it will accelerate in the direction of the negative mass particle. If the pair is bound gravitationally it will accelerate in the direction of the positive mass particle. Finally, Bondi's solution requires the existence of two such pairs with opposite orientation.

The models of mass dipoles provided by Blanchet, Blanchet and Le Tiec, Hajdukovic, and Penner, all lead to the BTFR. Unfortunately they all violate some aspect of current physical theory. This manuscript will not be proposing another model of a mass dipole. Instead it will look at the possibility of mass dipoles and the anti-screening of a baryonic mass from a different perspective. It will start by considering the case where the polarization of the vacuum does occur, namely in electrodynamics. A semiclassical model of the screening of electric charge by virtual electric dipoles will be presented. This model is then applied to the hypothetical case of an electric force where like charges attract. The resulting anti-screening of the electric charge is found to have the same functional dependence on the field source and observation distance that is found with the Baryonic Tully-Fisher Relationship. This leads to an anti-screening model for the gravitational force which is then used to determine the theoretical rotational curve of the Galaxy and the theoretical velocity dispersions and shear values for the Coma cluster. These theoretical results are found to be in good agreement with the corresponding astronomical observations.

## 2. Theory

### 2.1 Model of electrodynamic screening

In quantum electrodynamics a charged particle is taken to be surrounded by a cloud of virtual photons. These photons spend part of their existence dissociated into pairs of virtual particle-antiparticles of equal and opposite charge. The energy required to create each virtual particle, borrowed from the vacuum, as well as the time duration of their existence falls out from the Heisenberg uncertainty relation. While the virtual pair is in existence the virtual particle with charge opposite to that of the particle will, on average, be closer to the particle than the virtual particle of like sign and in effect the vacuum behaves as a dielectric medium. As with the classical model of a point charge within a dielectric, the virtual electric dipole pairs within the vacuum provide a screening effect so that at large observational distances the apparent charge of the particle is reduced. However, as the charged particle is approached the screening effect is reduced thereby resulting in more of the charge of the particle being observed. The apparent charge of a particle such as the electron is therefore actually smaller than its true non-screened value.

To model this behavior of a charged particle surrounded by a sea of virtual electric dipoles, a semiclassical model based on a point charge within a dielectric will be used. In this semiclassical model the virtual electric dipole moment density, $\mathbf{P_E}$, within the vacuum will be taken to be given by

$$\mathbf{P_E} = n\mathbf{p} \tag{3}$$

where $\mathbf{p}$ is the average electric dipole moment of the virtual pairs and n is the density of the virtual pairs. The value of $\mathbf{p}$ will be determined not only by the charge and the average separation of the virtual pair but also by their lifetime. The value of n would be expected to be proportional to both the density of the virtual photons and to the probability of a given virtual photon being dissociated into a virtual pair. The resulting volume charge density of the vacuum, $\rho_v$, surrounding a given particle will then be given by

$$\rho_v = -\nabla \cdot \mathbf{P_E} . \tag{4}$$

The resulting electric field contribution, $\mathbf{E_V}$, due to this polarized vacuum will in turn be given by

$$\mathbf{E_V} = \frac{1}{4\pi\epsilon_o} \int_V \frac{\rho_v \hat{r}}{r^2} dV. \tag{5}$$

In the case of spherical symmetry or in the far field limit, where $r \to \infty$, (5) with the use of (4) simplifies to

$$\mathbf{E_V} = -\frac{1}{\epsilon_o} \mathbf{P_E}. \tag{6}$$

In the classical dielectric model a surface charge, $\sigma_v$, will also exist on a cavity which is taken to surround the point charge. The total electric field, $\mathbf{E}$, surrounding a particle of charge $Q_o$ will therefore be given by

$$\mathbf{E} = \mathbf{E_Q} + \mathbf{E_V} + \mathbf{E_S} . \tag{7}$$

Here $\mathbf{E_Q}$ is the electric field due directly to the particle and $\mathbf{E_S}$ is the electric field due to the surface charge. In general, both $\rho_v$ and $\sigma_v$ will be of opposite sign to $Q_o$ thereby resulting in the vacuum contribution reducing the total electric field. This reduction in the total electric field is equivalent to a reduction of the apparent charge.

In this model it will be taken that the dipole moment density depends on the strength of the electric field. First, the case where the dependence of the dipole moment density on the total electric field is linear will be considered;

$$\mathbf{P_E} = n\mathbf{p} = \chi \epsilon_o \mathbf{E} \tag{8}$$

where $\chi$ is a dimensionless constant known in the classical case as the electric susceptibility of a dielectric. This linear relationship fundamentally corresponds to the density of the virtual dipole pairs being directly proportional to the density of the virtual photons which in turn is directly proportional to the magnitude of the electric field. Classically, (8) leads to the volume charge density being equal to zero throughout the dielectric and to a surface charge, $Q_S$, on the surrounding cavity of (Lorrain & Corson 1970)

$$Q_S = -\left(\frac{\chi}{1+\chi}\right) Q_o. \tag{9}$$

The resulting total electric field will therefore be that of a point charge of magnitude

$$Q = Q_o + Q_S = \left(\frac{1}{1+\chi}\right) Q_o . \tag{10}$$

The apparent charge is therefore less than the true charge. Although, this model does lead to the apparent charge of the particle being reduced it does not lead to a reduction in the screening as the charged particle is approached, i.e. the apparent charge of the particle is observed to be independent of the observation distance.

The reason for there being no reduction in the screening effect is that (8) is taken to continue to apply even as r→ 0 or as E → ∞. In order for this model to lead to a decrease in the screening, as E is increased $\mathbf{P_E}$ must fall off of the linear relationship that is given by (8). The virtual dipole moment density may still increase as the charge is approached but not as fast as being proportional to $\mathbf{E}$, i.e. the density of the virtual photons. To get the proper behavior therefore requires

$$\mathbf{P_E} \to \chi \epsilon_o \mathbf{E} \quad \text{as } E \to 0 \quad \text{and} \tag{11a}$$

$$\mathbf{P_E} < \chi \epsilon_o \mathbf{E} \quad \text{as } E \to \infty . \tag{11b}$$

The specific function chosen to model this behavior is found to have only a secondary effect. For our purposes, the following function, which has the behavior given by (11), will be used

$$\mathbf{P_E} = \chi \epsilon_o E_o \left(1 - e^{-E/E_o}\right) \hat{\mathbf{E}} \tag{12}$$

where $E_o$ is a constant to be determined. For this function the value of $\mathbf{P_E}$ actually saturates as E → ∞ although this is not a requirement for the model.

For the dependence of $\mathbf{P_E}$ on E as given by (12) the surface charge $Q_S \to 0$ and the volume charge density is found to be no longer equal to zero. From (6), (7), and (12) the magnitude of the total electric field surrounding a point charge will in this case be given by

$$E = E_Q - \chi\, E_o \left(1 - e^{-E/E_o}\right). \tag{13}$$

Substituting

$$E_Q = \frac{1}{4\pi\epsilon_o} \frac{Q_o}{r^2} \quad \text{and} \tag{14}$$

$$E = \frac{1}{4\pi\epsilon_o} \frac{Q}{r^2} \tag{15}$$

into (13) leads to

$$Q = Q_o - 4\pi\epsilon_o E_o \chi\, r^2 \left(1 - e^{-Q/(4\pi\epsilon_o E_o r^2)}\right). \tag{16}$$

Using the following change of variables; $Q' = Q/4\pi\epsilon_o E_o r^2$, $Q_o' = Q_o/4\pi\epsilon_o E_o r^2$ and $u = Q' - Q_o' + \chi$, then leads to

$$u e^u = \chi\, e^{-(Q_o' - \chi)} \tag{17}$$

which has the solution

$$u = W\!\left(\chi\, e^{-(Q_o' - \chi)}\right) \tag{18}$$

where $W(x)$ is the Lambert W function. Substituting back into (18) then results in the following relationship between Q, the apparent charge, $Q_o$, the true charge, r, the observation distance, and $\chi$ the susceptibility;

$$Q = Q_o - 4\pi\epsilon_o E_o \chi\, r^2 + 4\pi\epsilon_o E_o r^2\, W\!\left(\chi e^{-(Q_o - 4\pi\epsilon_o E_o \chi\, r^2)/(4\pi\epsilon_o E_o r^2)}\right). \tag{19}$$

Setting the constant $E_o$ to

$$E_O = \frac{1}{4\pi\epsilon_o} \frac{e}{r_o^2} \tag{20}$$

where *e* is the charge of an electron and $r_o$ is a reference distance to be determined then leads to

$$Q/e = Q_o/e - \chi \left(\frac{r}{r_o}\right)^2 + \left(\frac{r}{r_o}\right)^2 W\!\left(\chi e^{-((Q_o/e) - \chi(r/r_o)^2)/(r/r_o)^2}\right). \tag{21}$$

    To model specifically the screening of an electron, the true charge $Q_o$ will be set equal to $(1+\chi)e$ in (21) so that the apparent charge will equal *e*. Figure 1 shows the resulting dependence that the apparent charge of the electron has on observation distance for various values of $\chi$. As is seen, the magnitude of the apparent charge drops from $(1+\chi)e$ to *e* over a distance of $\sim r_o$ from the electron. The model of the screening that leads to Figure 1 is in qualitative agreement with the general results found in QED. As a semiclassical model, it does have limitations. However, semiclassical or not, in order to have the apparent charge at large distances being constant and increasing to its true charge value as it is approached, as shown in Figure 1, requires the behavior of the dipole moment density as given by (11). Again the particular function used, i.e. (12), to model this dependence is secondary.

## 2.2 Model of electrodynamic anti-screening

Consider now the hypothetical case where the electric force is such that like charges attract and unlike charges repel. In this case, the virtual particle with the same charge as that of the particle will, on average, be closer to the particle than the virtual particle of unlike sign. The vacuum volume charge density $\rho_v$ will now be of the same sign as that of the particle. As such, the total electric field will be increased and the apparent charge will be greater than its true value. To apply the model of Section 2.1 to this anti-screening just requires negative values for $\chi$ to be used. However, although the solution given by (21) is applicable to negative values of $\chi$, the Lambert W function has two branches for negative arguments. These branches are labeled $W_0$ for the upper branch and $W_{-1}$ for the lower branch. Setting the condition that $Q \to Q_o$ as $(r/r_o) \to 0$, which corresponds from (21) to $(r/r_o)^2 W(f(r/r_o) \to 0^-) \to 0$ as $(r/r_o) \to 0$, selects the upper branch as $(r/r_o)^2 W_{-1}(f(r/r_o) \to 0^-) \to -1$ as $(r/r_o) \to 0$. By the use of (21), Figure 2 shows the dependence that the apparent charge of the electron has on observation distance for various negative values of $\chi$. For this case the value of the true charge $Q_o$ has been set to $e$. As seen in the figure, the apparent charge increases with the observation distance. Of particular interest is the case where $\chi = -1$ where at large observational distances the apparent charge depends linearly on observation distance. This is the same behavior that is found with the BTFR. Figure 3 shows the corresponding vacuum volume charge density profiles where the value of $r_o$ has been set equal to the Compton wavelength for an electron. For $\chi = -1$ the profile is found to fall off as $1/r^2$. Figure 4 shows the relationship between apparent charge and true charge for $\chi = -1$ with the observation distance $r/r_o$ set equal to 10 in (21). At this large observation distance, the apparent charge is found to be approximately proportional to the square root of the true charge. This is again in agreement with the functional dependence found with the BTFR.

This dependence that the apparent charge has on observation distance and the value of the true charge can also be found by considering the far field limit of (16) with $\chi = -1$. To 2nd order, where $r/r_0 \gg 1$, (16) becomes

$$Q = Q_o + Q - \frac{Q^2}{8\pi\epsilon_o E_o r^2} \ . \tag{22}$$

which simplifies to

$$Q = (8\pi\epsilon_o E_o)^{1/2} \, r \, Q_o^{1/2}. \tag{23}$$

This analysis shows that the same dependence of $\mathbf{P_E}$ on field strength, i.e. (12), leads to two very different outcomes for the two possible types of long range forces. For forces where unlike charges attract, the screening effect reduces the value of the apparent charge with the apparent charge for large observational distances being constant. For forces where like charges attract, and in the case where $\chi = -1$, one has anti-screening where at large observational distances the apparent charge has the dependence on the observation distance and the true charge as given by (23). Again the particular function used for $\mathbf{P_E}$ in the model plays a secondary role and in the case of the anti-screening will only effect the coefficient in (23).

## 2.3 Model of gravitational anti-screening

The agreement between (23) and the BTFR, i.e. (2), is quite remarkable. The screening of electric charge as found with QED and the larger apparent mass of galaxies and clusters would appear to be two sides of the same coin. Carrying this connection forward leads to the following hypothesis:

> Particles and thereby baryonic masses are surrounded by a cloud of virtual gravitons which spend part of their existence dissociated into virtual entities. The virtual entities have no net mass but do have a mass dipole moment. In addition, the dependence that the virtual mass dipole moment density has on the gravitational field is the same as the modeled dependence, as given in Section 2.2, that the virtual electric dipole moment density has on the electric field.

Whether the hypothesized entities are related to any of the problematic mass dipole models discussed in Section 1 or are something completely different will not be speculated on.

Given the above hypothesis the results of Section 2.2 can easily be carried over to the gravitational case. The only modifications to the model of anti-screening required are basically with respect to the constants, i.e. replacing $\epsilon_o$ with $1/4\pi G$. Equations (11a) and (11b) thereby become

$$\mathbf{P_G} \rightarrow \chi \frac{1}{4\pi G} \mathbf{g} \quad \text{as } g \rightarrow 0 \quad \text{and} \tag{24a}$$

$$\mathbf{P_G} < \chi \frac{1}{4\pi G} \mathbf{g} \quad \text{as } g \rightarrow \infty \tag{24b}$$

with $\mathbf{P_G}$ being the virtual mass dipole moment density and g being the total gravitational field. Equations (12), (13) and (16) become respectively

$$\mathbf{P_G} = \chi \frac{1}{4\pi G} g_o (1 - e^{-g/g_o}) \hat{\mathbf{g}}, \tag{25}$$

$$g = g_M - \chi g_o (1 - e^{-g/g_o}) \quad \text{and} \tag{26}$$

$$M = M_o - \frac{g_o}{G} \chi r^2 \left(1 - e^{-GM/g_o r^2}\right). \tag{27}$$

where $g_M$ is the gravitational field due to the baryonic mass and $g_o$ is a constant to be determined. The relationship between M, the apparent mass, $M_o$, the baryonic mass, r, the observation distance, and $\chi$, the susceptibility follows from solving (27) or making the appropriate substitutions in (19);

$$M = M_o - \frac{g_o}{G} \chi r^2 + \frac{g_o}{G} r^2 W\left(\chi e^{-(GM_o - \chi g_o r^2)/g_o r^2}\right). \tag{28}$$

Setting $\chi = -1$ in (27) leads to the same far field dependence for the gravitation field that was found in the electrodynamic case, i.e. (23);

$$M = \left(\frac{2g_o}{G}\right)^{1/2} r\, M_o^{1/2}, \tag{29}$$

which of course is the BTFR. In this case we can obtain the value of the constant $g_o$ from observations. Equating the coefficients of (2) and (29) leads to the following value for $g_o$;

$$g_o = \frac{1}{2GA} = (8.0 \pm 1.0) \times 10^{-11} \text{ m s}^{-2}. \tag{30}$$

Although the prime motivation behind this theory of gravitational anti-screening is the agreement between (29) and the BTFR, the theory does more than just provide a natural explanation for the BTFR. To show the theory in action it will now be applied to the two primary observations where dark matter is evoked.

## 3. Applications of the theory

### 3.1 Extended baryonic mass distributions

In general, to determine the contribution of the polarized vacuum to the gravitational field for a baryonic mass distribution requires (25), with $\chi$ set equal to -1, along with the following gravitational equivalents of (4) and (5);

$$\rho_v = -\nabla \cdot \mathbf{P_G} \quad \text{and} \tag{31}$$

$$\mathbf{g_v} = G \int_V \frac{\rho_v \hat{r}}{r^2} dV \tag{32}$$

where $g_v$ is the gravitational field due to the polarized vacuum. An iterative technique is used where the initial estimate of the total gravitational field is taken to be solely that due to the baryonic mass. The value of $\mathbf{P_G}$ and the resulting equivalent mass density distribution of the vacuum is then determined by (25) and (31) along with the value of $g_o$ from (30). From the mass density distribution the gravitational field due to the vacuum is determined from (32) and for the next estimate the total gravitational field is taken to be equal to the sum of the fields due to the baryonic mass and the vacuum. This iterative process is repeated until the resulting values of the total gravitational field obtained after a given iteration vary by less than 1% from the previous iteration.

The results presented in the following Sections 3.2 and 3.3 are a rework of some the authors previous results (Penner 2013a, Penner 2013b, Penner 2013c). For the results presented here the new $\mathbf{P_G}$, i.e. (25), was used.

### 3.2 Galactic rotation curves

As an application of the anti-screening theory the rotational curve of our own galaxy will be generated. As a first step the baryonic mass distribution of the Galaxy is required. Flynn et al (2006) provide estimates of the stellar mass of the Galaxy based on measurements of the volume luminosity density and surface luminosity density generated by the local Galactic disk. Using their results with the exponential disc scalelength set to 3 kpc, along with the approximation that the distribution of the gas follows that of the stellar mass, leads to the following baryonic mass components of the Galaxy:

$$M_{DISC} = (40 \pm 3) \times 10^9 \text{ M}_\odot \tag{33a}$$

$$M_{BULGE} = (21 \pm 2) \times 10^9 \text{ M}_\odot. \tag{33b}$$

The specific distributions of these baryonic mass components are provided for in Penner (2013a).

Following the procedure described in Section 3.1, the resulting vacuum mass density profile for the Galaxy was generated. As the baryonic distribution is not spherically symmetric neither is the mass density distribution for the polarized vacuum. As such, Figure 5 shows the mass density profile of the vacuum along both the plane of the Galaxy and perpendicular to the plane. For large r both distributions fall off as $1/r^2$ as expected. From the generated gravitation field, in the plane of the galaxy, the theoretical rotational curve for the Galaxy was then determined. This is shown on Figure 6 along with the contributions from the baryonic mass components and the polarized vacuum. As is seen, the polarized vacuum contribution starts to dominate for r > 10 kpc. The theoretical rotational velocity at 8 kpc (the position of the Sun) is found to be 223 km s$^{-1}$. The galactic rotational velocity is then found to slowly fall from this value as it asymptotically approaches a value of 190 km s$^{-1}$. This corresponds to the BTFR value. Included on Figure 6 is the compilation of results as provided by Sofue (2012) scaled to the standard value of 220 km s$^{-1}$ at the Sun's location. On Figure 6 there is good agreement between the anti-screening theory and the observations. It needs to be stressed that there are no free parameters being used. The only parameter in the theory is the value of A as given by (1b) which is determined observationally. The rotational curve of the Galaxy as shown in Figure 6 is determined by its baryonic mass distribution. The specific function used for **P**$_G$ does play a secondary role however. For the author's previous model, which used a different **P**$_G$, the rotational value at the location of the Sun was found to be 235 km s$^{-1}$ (Penner 2013a).

A major difference between the rotational curves obtained from the anti-screening model presented here and those obtained from the dark matter model is that the theoretical anti-screening rotation curves depend on the details of the baryonic mass distribution. As an example, on Figure 7, theoretical rotational curves are shown for galaxies of mass 60 x 10$^9$ M⊙ whose baryonic mass distribution parameters are the median values of three classes of spiral galaxies i.e. Hubble stages T=1, 3, and 5. The values of the parameters are as given in Penner (2013b). As is seen the behavior of the inner parts of the rotation curve depends strongly on the details of the baryonic mass distribution. Other examples of how the baryonic mass distribution affects the rotational velocity curve are also provided in Penner (2013b). Although, the specific **P**$_G$ used to generate these rotational curves is different, the effect of the difference is only secondary. Overall, from Figure 7 and these previous results, it is found that the rotational curves generated by the anti-screening theory readily produces features observed in the rotational curves of spiral galaxies. Indeed, unlike the dark matter model, given the astronomically determined baryonic mass distribution of a galaxy the rotation curve for the galaxy can be predicted.

## 3.2 Coma cluster

As an example of applying the anti-screening theory to a cluster, the Coma cluster will be considered. In this case, there are three major components of the baryonic mass. The intra-cluster gas (ICG) is the dominant baryonic component. Observations of the X-ray emissions find the ICG distribution to be approximately spherically symmetric about the cluster centre. If the gas is taken to be isothermal, the gas density profile is given by

$$\rho_{gas} = \rho_{go}\left(1 + \left(\frac{r}{r_c}\right)^2\right)^{-\frac{3}{2}\beta_{gas}}. \tag{34}$$

with $r_c = 0.309$ Mpc, $\beta_{gas} = 0.75$, and the central gas density $\rho_{go} = 6.49 \times 10^{-24}$ kg m$^{-3}$ (Briel et al 1992, Lokas and Mamon 2003). If this profile is taken to be valid out to 3 Mpc, corresponding to the radial extent that X-ray emission has been detected, the total ICG mass within 3 Mpc is $2.0 \times 10^{14}$ M$_\odot$. Also considered will be the case where the ICG is more concentrated with $\beta_{gas}$ in (34) being 50% greater but with $\rho_{go}$ adjusted so that there is the same amount of ICG within 3 Mpc as with the isothermal profile. (Penner 2013c).

After the ICG, the second most important contributor to the baryonic mass of the Coma cluster are the stars within the cluster's galaxies. Lokas&Mamon (2003) fitted a NFW profile to the surface luminosity data for the Coma cluster. Their estimated mean mass-to-light ratio of 6.43 M$_\odot$/L$_\odot$ leads to a total stellar mass of $2.6 \times 10^{13}$ M$_\odot$ within 3 Mpc and a total stellar mass of $3.9 \times 10^{13}$ M$_\odot$ within 6 Mpc. The total stellar mass of the galaxies is therefore approximately one order of magnitude less than the estimated ICG mass.

The third contributor to the baryonic mass of the cluster is the gas localized around the two giant elliptical galaxies, NGC 4874 and NGC 4889, which are found near the centre of the cluster. From measurements of the sub clustering around the two ellipticals (Mellier et al 1988) and their X-ray luminosity (Vikhlinin et al 1994) it is estimated that the total amount of gas concentrated around NGC 4874 and NGC 4889 is $1.5 \times 10^{13}$ M$_\odot$. A model of the distribution of this gas is provided in Penner (2013c).

Following the procedure described in Section 3.1, the resulting vacuum mass density profiles for the Coma cluster for both ICG distributions are shown on Figure 8. As the baryonic mass distribution is spherically symmetric so is the vacuum mass density. At large r the anti-screening mass density distributions again fall off as $1/r^2$. The corresponding theoretical velocity dispersions for the galaxies of the Coma cluster were calculated from the generated gravitational field. These are shown on Figure 9 in the case of the galaxies being in both radial and isotropic orbits. Details of the derivation of these velocity dispersions are given in Penner (2013c). Included on this figure are the observed velocity dispersions as given by Lokas&Mamon (2003) and Merritt (1987). The theoretical results very clearly predict that the outer galaxies in the Coma cluster are in radial orbits. Dark matter models with their free parameters cannot make such clear predictions. Theoretical shear values for the Coma cluster generated by the theory for the two ICG distributions are in turn shown on Figure 10. The derivation of these can again be found in Penner (2013c). Included on this figure are the measured values as given by Kubo et al (2007) and Gavazzi et al (2009). As Figures 9 and 10 show, the theory and model of anti-screening is in good agreement with the observations of the Coma cluster. Comparing these results to previous results (Penner 2013c) it is again found that the specific **P$_G$** has little effect.

Again, it needs to be stressed that, as with the galactic rotational curves, there are no free parameters involved. Although, the uncertainties with regard to the gas distributions do allow for a certain amount of fitting, the only true parameter in the model, A, is determined by the BTFR. This coefficient and the baryonic mass distribution are the only inputs required in the determination of the dynamics of both galaxies and clusters. Also, the outer galaxies in the Coma cluster being in radial orbits falls out from the theory. Therefore, unlike the dark matter theory, predictions based off of the baryonic mass distribution can be made.

## 4. Conclusion

It has been shown that from a model of the screening effect produced by a virtual electric dipole field surrounding a charged particle, a model of the anti-screening of a baryonic mass by a virtual gravitational mass dipole field can be obtained. The anti-screening in the case where $\chi = -1$ was found to have the exact same dependence on the baryonic mass and the observation distance as that found with the Baryonic Tully-Fisher relationship. The correspondence between (2) and (29) may be coincidence but the author is of the opinion that it points towards the solution of the dark matter problem.

Good agreement was then found between the theoretically generated rotational curve of the Galaxy, the theoretical generated velocity dispersions and shear values for the Coma cluster, and astronomical observations. The positive results for the Coma cluster are especially encouraging. The equation for $\mathbf{P}_G$ used for the cluster is exactly the same as that used for the Galaxy. The only difference between modeling the dynamics of galaxies and the dynamics of clusters is their baryonic mass distribution, which can be obtained from observations.

The model of virtual gravitons dissociating into virtual entities that have a mass dipole moment should be seen as a first step to understanding how the vacuum becomes polarized. It does seem to indicate that a future theory of quantum gravity will lead to an explanation the observations currently attributed to dark matter. The direction of future research for the author will be directed towards applying the anti-screening theory and model to other observations where dark matter is evoked. It will be demonstrated that not only does the theory provide a better explanation of these observations but it has a predictive power that the dark matter model does not.

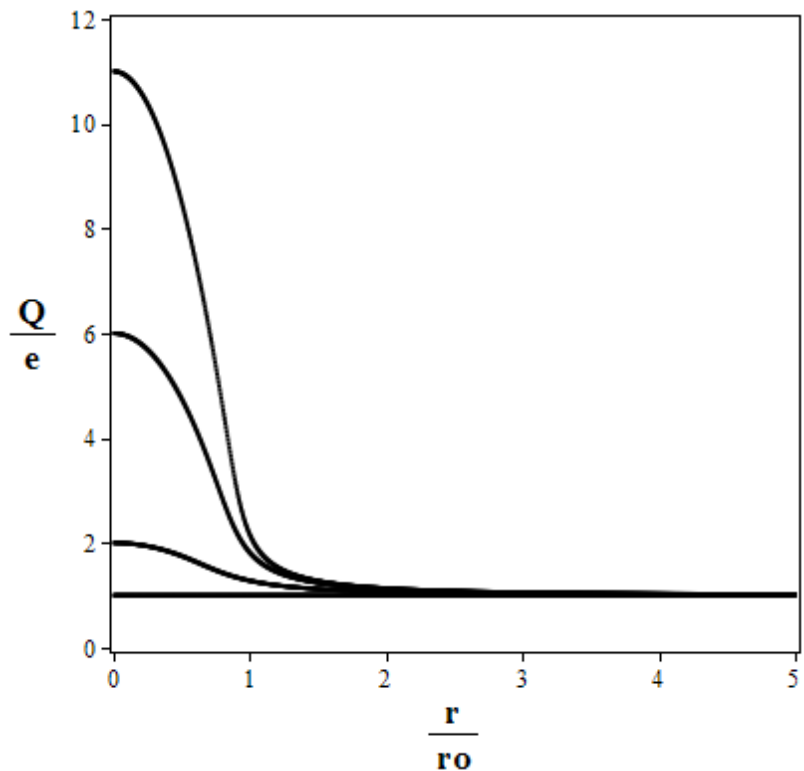

**Figure 1**: The apparent charge as a function of observation distance in the case of screening with $Q_o = (1+ \chi)$ e for $\chi = 0, 1, 5$ and $10$ (in order from lower curve to upper curve).

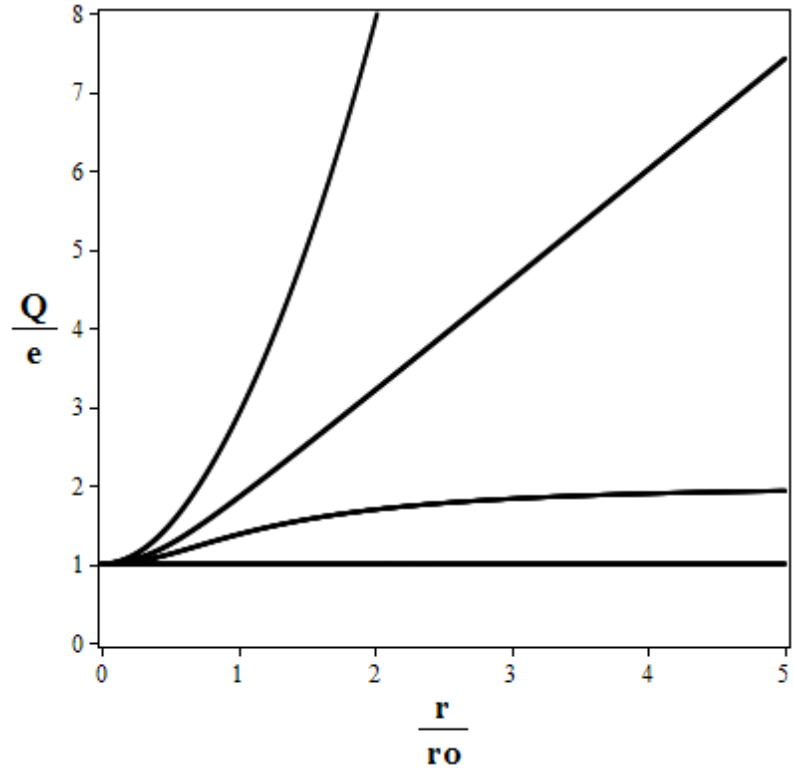

**Figure 2**: The apparent charge as a function of observation distance in the case of anti-screening with $Q_o = e$ for $\chi = 0$, -0.5, -1 and -2 (in order from lower curve to upper curve).

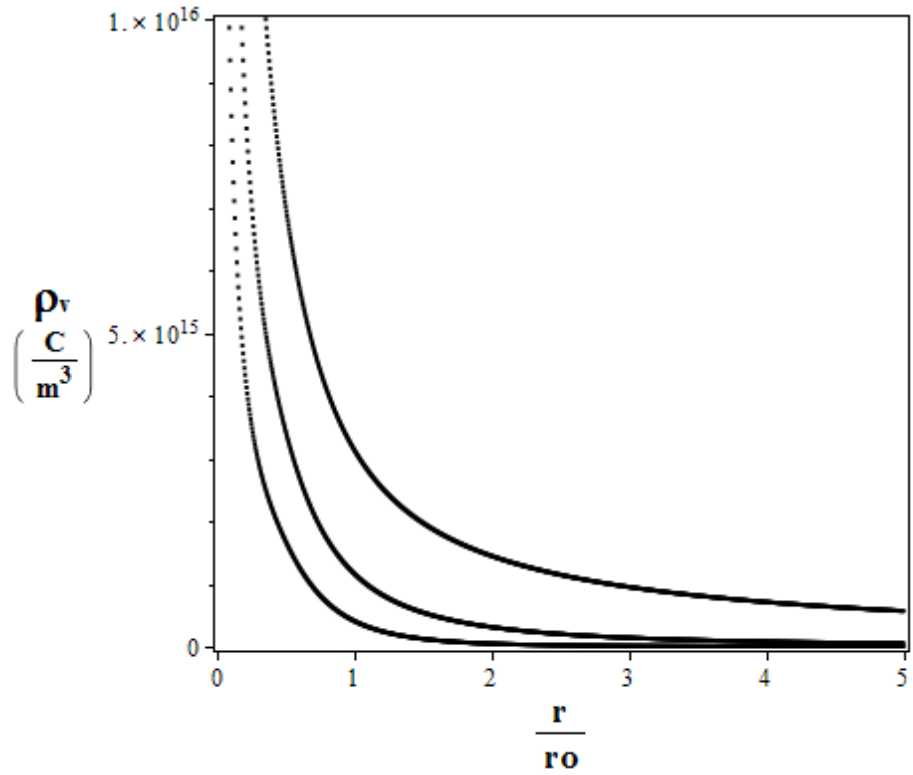

**Figure 3**: The vacuum charge density profile in the case of anti-screening with $Q_o = e$, $r_o = h/m_e c$ for $\chi = -0.5, -1, -2$ (in order from lower curve to upper curve).

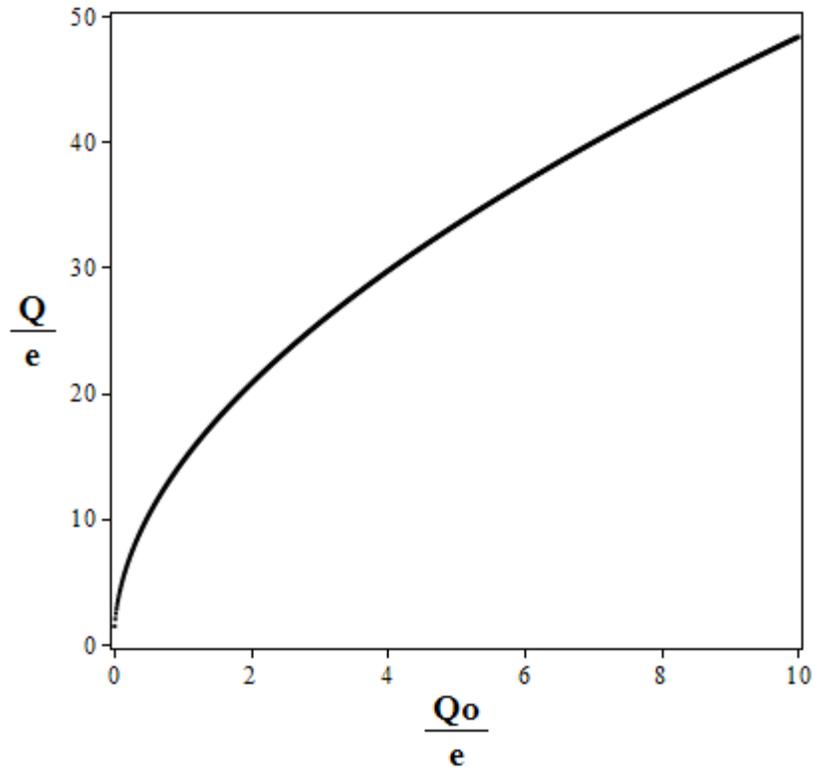

**Figure 4**: The dependence of the apparent charge on the true charge in the case of anti-screening with $Q_o = e$, $\chi = -1$ and $r/r_o = 10$.

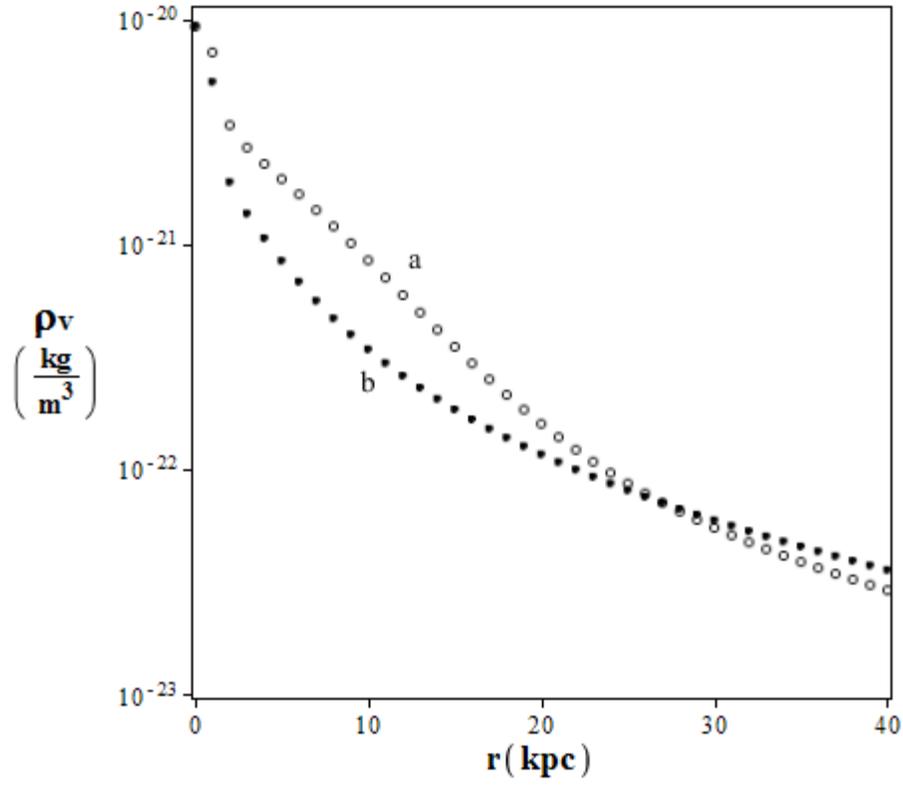

**Figure 5:** The vacuum mass density profile a) along the plane of the Galaxy and b) perpendicular to the plane.

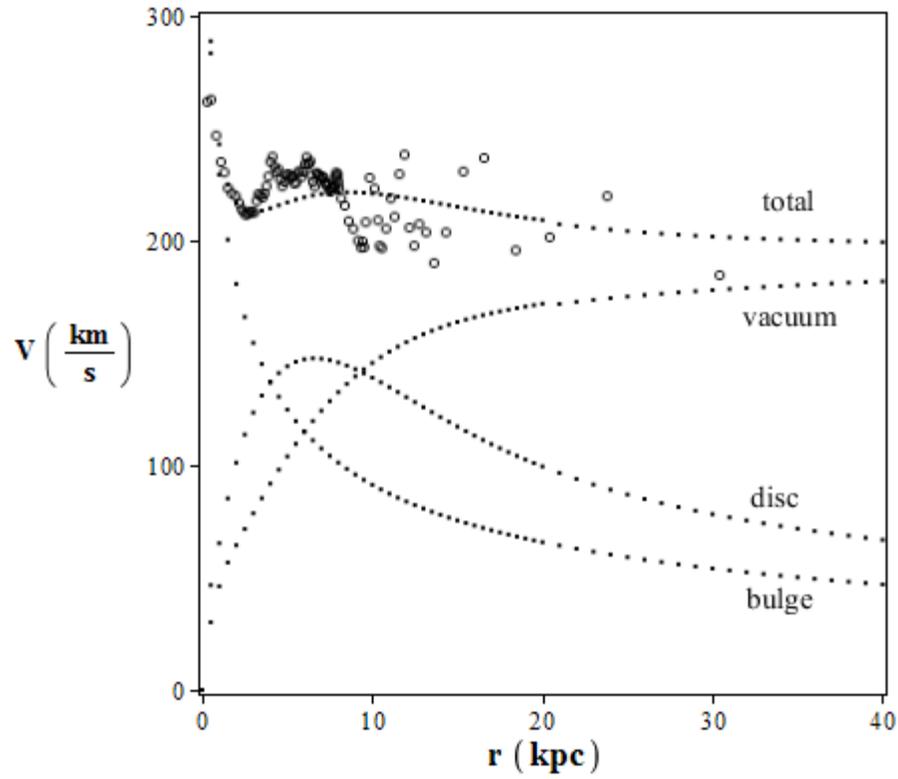

**Figure 6:** The theoretical contributions that the bulge, disc, and the polarized vacuum make to the total rotational curve of the Galaxy. Also shown (o) are the compilation of observations as given by Sofue (2012).

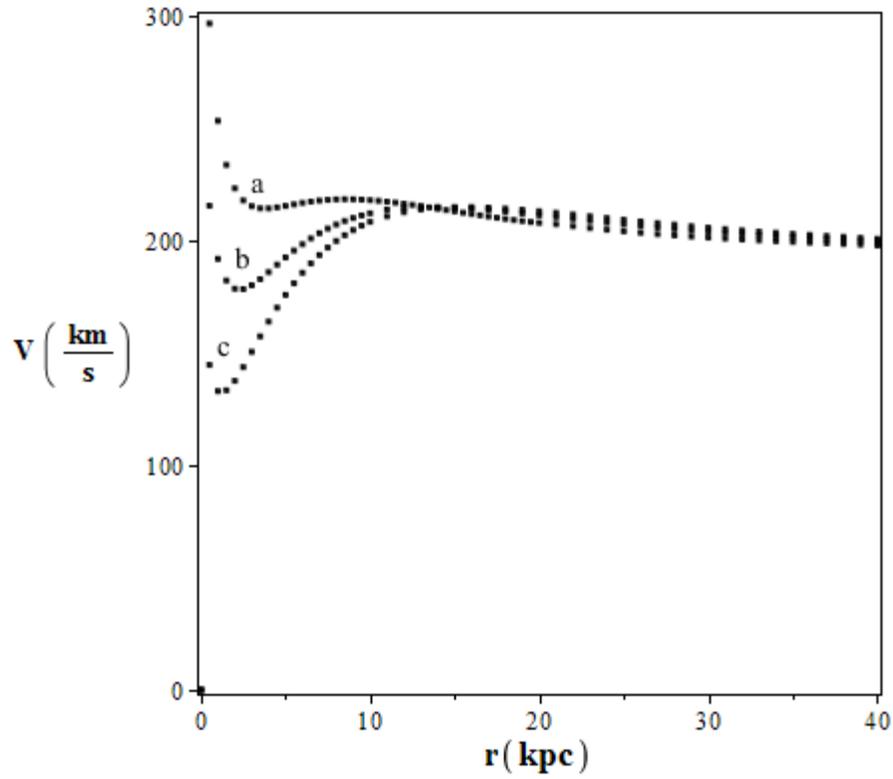

**Figure 7:** Theoretical rotational curves for modeled galaxies of mass 60 x $10^9$ M☉, whose baryonic mass distribution parameters are the median values of three classes of spiral galaxies. a) stage T=1, b) stage T=3, and c) stage T=5.

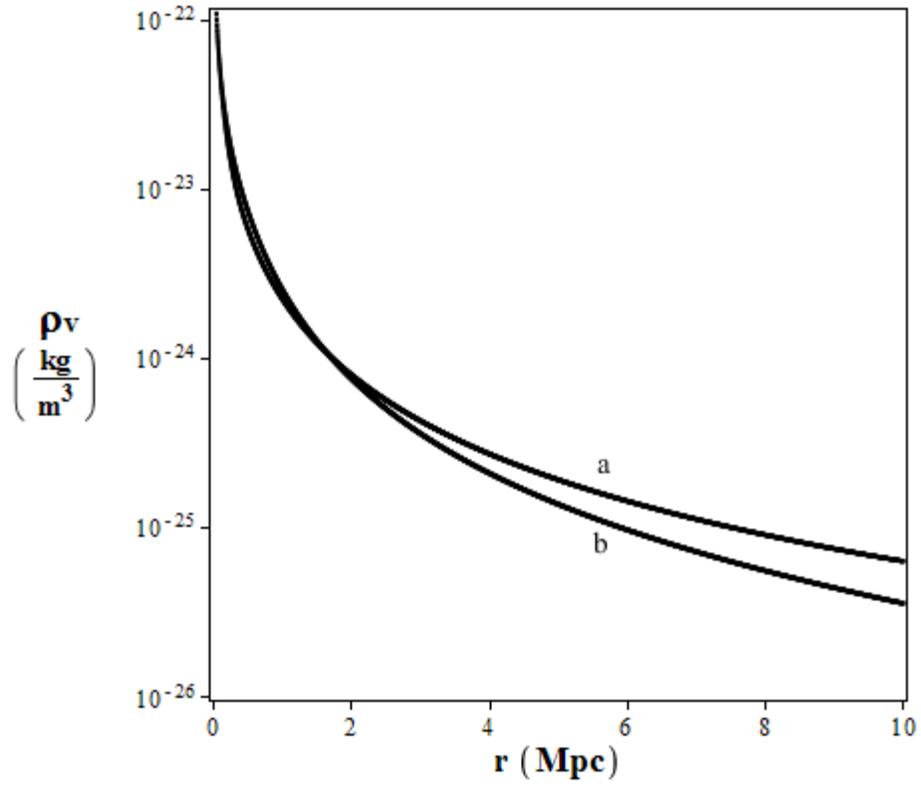

**Figure 8**: The vacuum mass density profile in the case of a) an isothermal intracluster gas and b) a more concentrated intracluster gas.

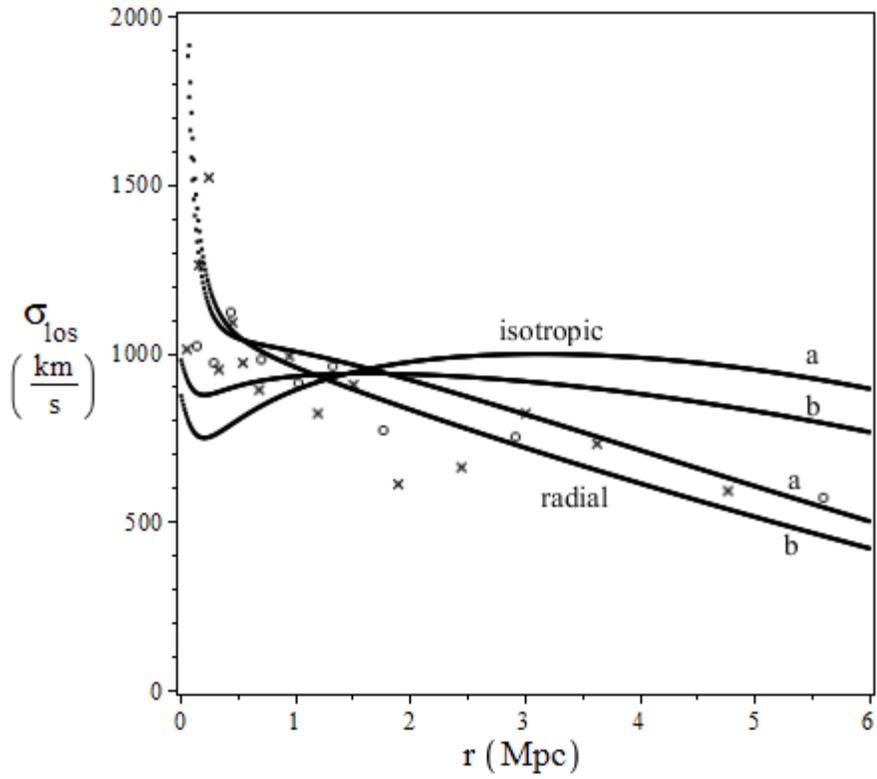

**Figure 9:** Theoretical velocity dispersions for the Coma cluster for radial and isotropic orbits in the cases of a) an isothermal intracluster gas and b) a more concentrated intracluster gas. Also included are the observed values as given by (o) - Lokas&Mamon (2003) and (x) - Merritt (1987).

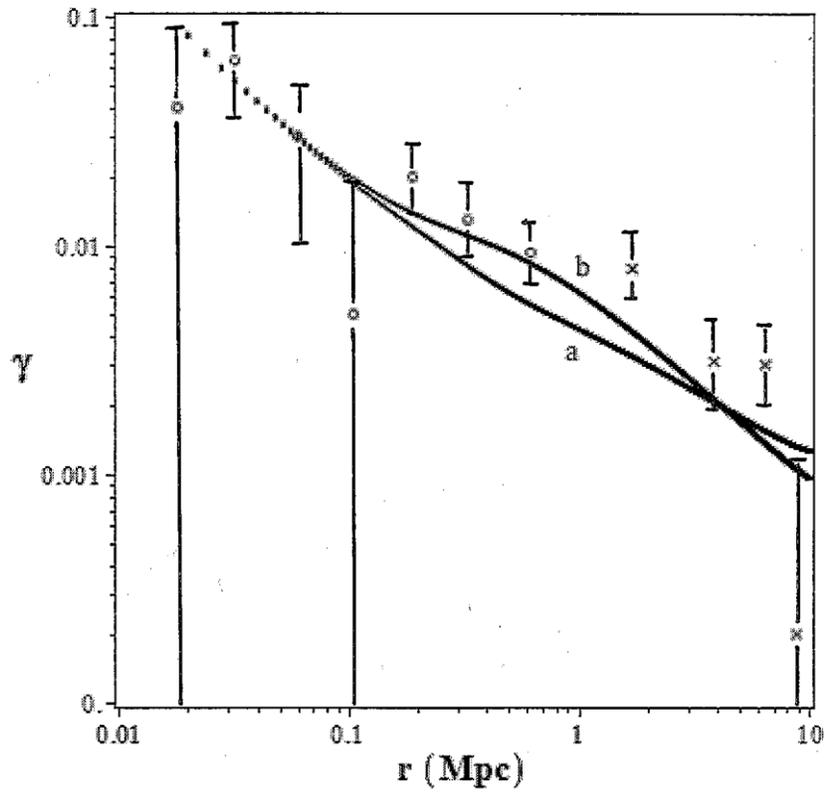

**Figure 10:** The theoretical shear values for the Coma cluster in the cases of a) an isothermal intracluster gas and b) a more concentrated intracluster gas. Also included are the shear measurements as given by (x) - Kubo et al (2007) and (o) - Gavazzi et al (2009).